\documentclass[slac_one]{revtex4}
\usepackage{graphicx}
\usepackage{fancyhdr}
\begin{document}
%Title of paper
\title{Top Quark Physics} %% Paper title goes here

% Repeat the \author .. \affiliation  etc. as needed
%
% \affiliation command applies to all authors since the last
% \affiliation command. The \affiliation command should follow the
% other information

\author{Erich W. Varnes for the CDF, D\O, and H1 Collaborations}
\affiliation{University of Arizona, Tucson, AZ 85721, USA}

\begin{abstract}
A review is presented of the current experimental status of the top quark sector of the standard model.  The measurements summarized include searches for electroweak single top production, the latest results on the $t\bar{t}$ production cross section, and searches for new physics in top quark production and decay.  In addition, the recent measurement of the top quark mass to a precision of 0.7\% is highlighted.

\end{abstract}

%\maketitle must follow title, authors, abstract
\maketitle

\thispagestyle{fancy}

% body of paper here - Use proper section commands
% References should be done using the \cite, \ref, and \label commands
% Put \label in argument of \section for cross-referencing
%\section{\label{}}

\section{INTRODUCTION} % Section title should be in all capitals.
The observation of the top quark in 1995~\cite{top_obs} provided an isospin partner for the $b$ quark, thereby completing the quark sector of the standard model (SM).  Within the context of the SM, every feature of the top quark is specified, aside from its mass.  In addition, assuming three quark generations, the CKM matrix element $|V_{tb}|$ is constrained to be very close to one~\cite{pdg}, meaning that the top quark decays nearly always to a $W$ boson and $b$ quark.    

In these proceedings, I review the current status of our knowledge of the top quark.  I start by discussing the search for electroweak single top production, and what we have learned about the top quark's couplings from that search. I then describe a series of measurements made using $t\bar{t}$ events, starting with the mass of the top quark, and continuing to its production and decay properties.  Aside from a single exception, these measurements are based on data collected by the D\O\ and CDF experiments at the Tevatron $p\bar{p}$ collider at Fermilab.  I close with a preview of the precision with which top quark properties are expected to be measured at the Large Hadron Collider.

\section{SINGLE TOP QUARK PRODUCTION}

Electroweak single top production is of interest because its rate depends directly on the strength of the top quark's coupling to a gauge boson.  Single top quarks can be produced via both s- and t-channel diagrams, and measuring both the total rate and the individual rates for t- and s-channel production probes the presence of contributions from non-SM couplings.  Searches for single top production make use of leptonic (electronic or muonic) decays of the $W$ boson from the top quark, which provides the signature of a high-$p_T$ lepton, missing $E_T$, and two or more jets (with two of the jets being $b$ jets). 

\subsection{Search for SM production}

At the Tevatron, the cross section for single top production is expected to be $\approx 3$pb. This is many orders of magnitude below the cross sections for the  $W$+jet and multijet processes that form the main backgrounds.  Selecting events with the jets, leptons, and missing $E_T$ consistent with single top production yields a sample with a signal-to-background ratio of $\approx 1/200$.  Extracting the single top contribution requires identification of $b$ jets and maximal use of the kinematic information  in the event.  The former is implemented using precision tracking to discriminate $b$ jets from promptly-decaying light-quark and gluon jets, while the latter relies on a set of multivariate classifiers.  These include classical likelihoods, neural networks, matrix element calculations and boosted decision trees.
To maximize sensitivity, the data is split into subsets based on the type of lepton, total number of jets, and number of $b$-tagged jets.  

With a data sample of 0.9 fb$^{-1}$, D\O\ combines the results from three multivariate analyses to extract a single-top cross production section of $4.7 \pm 1.3$ pb \cite{D0singletopxs}.  The significance of the signal corresponds to about 3.6 standard deviations (s.d.),  with a significance of 2.3 s.d expected for the SM.   Assuming that the excess above background is from single top production, a lower limit of 0.68 at 95\% C.L. is set on $|V_{tb}|$. CDF uses a 2.7 fb$^{-1}$ sample to perform four multivariate analyses ~\cite{CDFsingletopxs}.  The expected signal significance for these ranges from 3.8 s.d. to 5.0 s.d., and while all four analyses find evidence for single top production, the measured cross sections ($2.0^{+0.9}_{-0.8}$ --  $2.7^{+0.8}_{-0.7}$ pb) are smaller than the SM prediction, resulting in observed significances of 2.6 to 4.2 s.d..  While the evidence strongly favors electroweak single top production, neither CDF nor D\O\ can claim this definitively, a situation that should change as more data is analyzed.

\subsection{Search for anomalous couplings}

Non-SM top quark couplings can alter the rate of single top production (both in the s- and t-channels) and also the kinematic distributions of the decay products.  The lowest-dimensionality Lagrangian for the $tWb$ coupling has the form:

\begin{equation}
L= {g \over \sqrt{2}}W_\mu^-\bar{b}\gamma^\mu(f_1^LP_L+f_1^RP_R)t - {g \over \sqrt{2}M_W} \partial_\nu W_\mu^-\bar{b}\sigma^{\mu\nu}(f_2^LP_L+f_2^RP_R)t + h.c. 
\end{equation} 

\begin{figure*}[t]
\centering
\includegraphics[width=55mm]{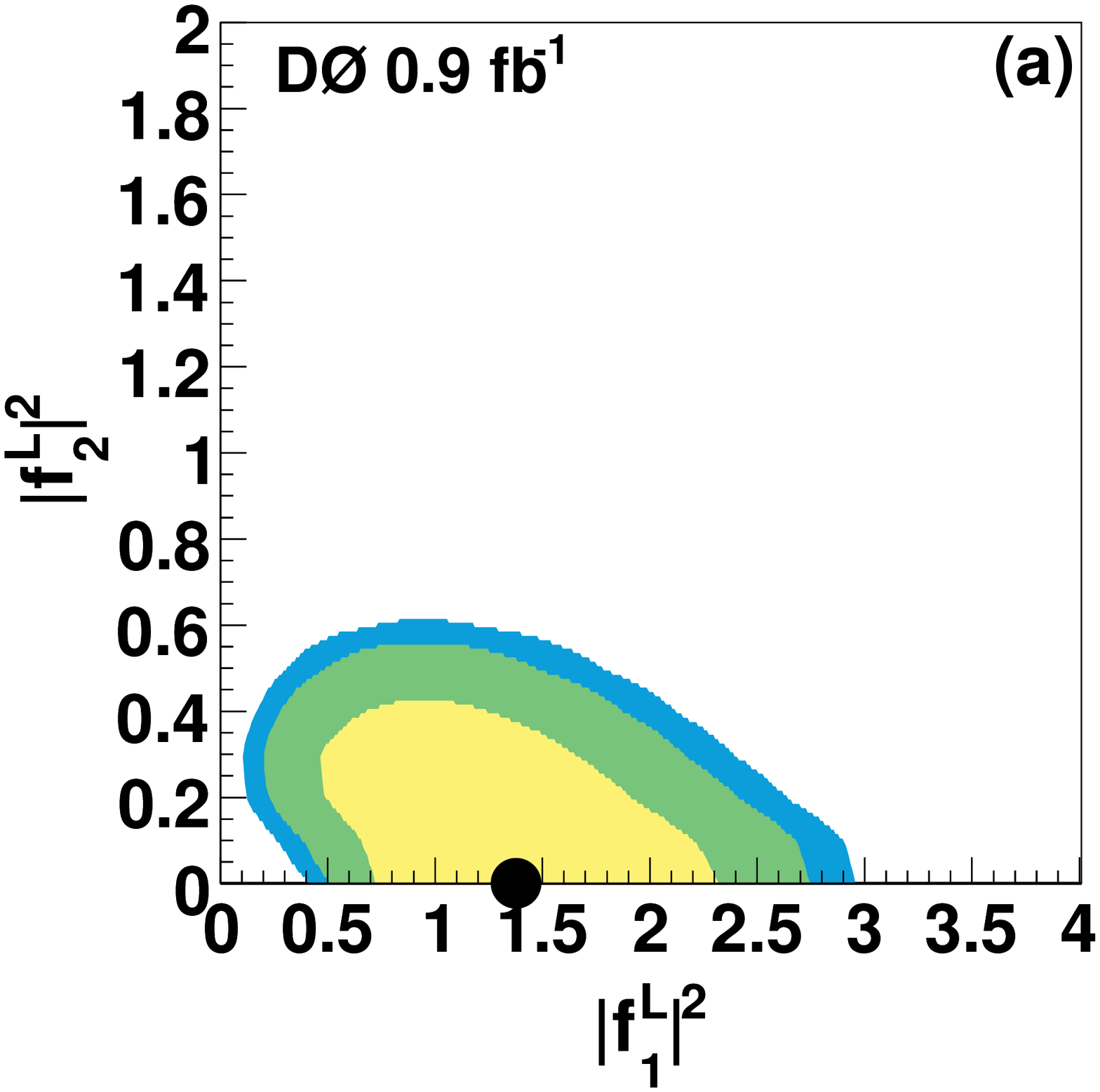}
\includegraphics[width=55mm]{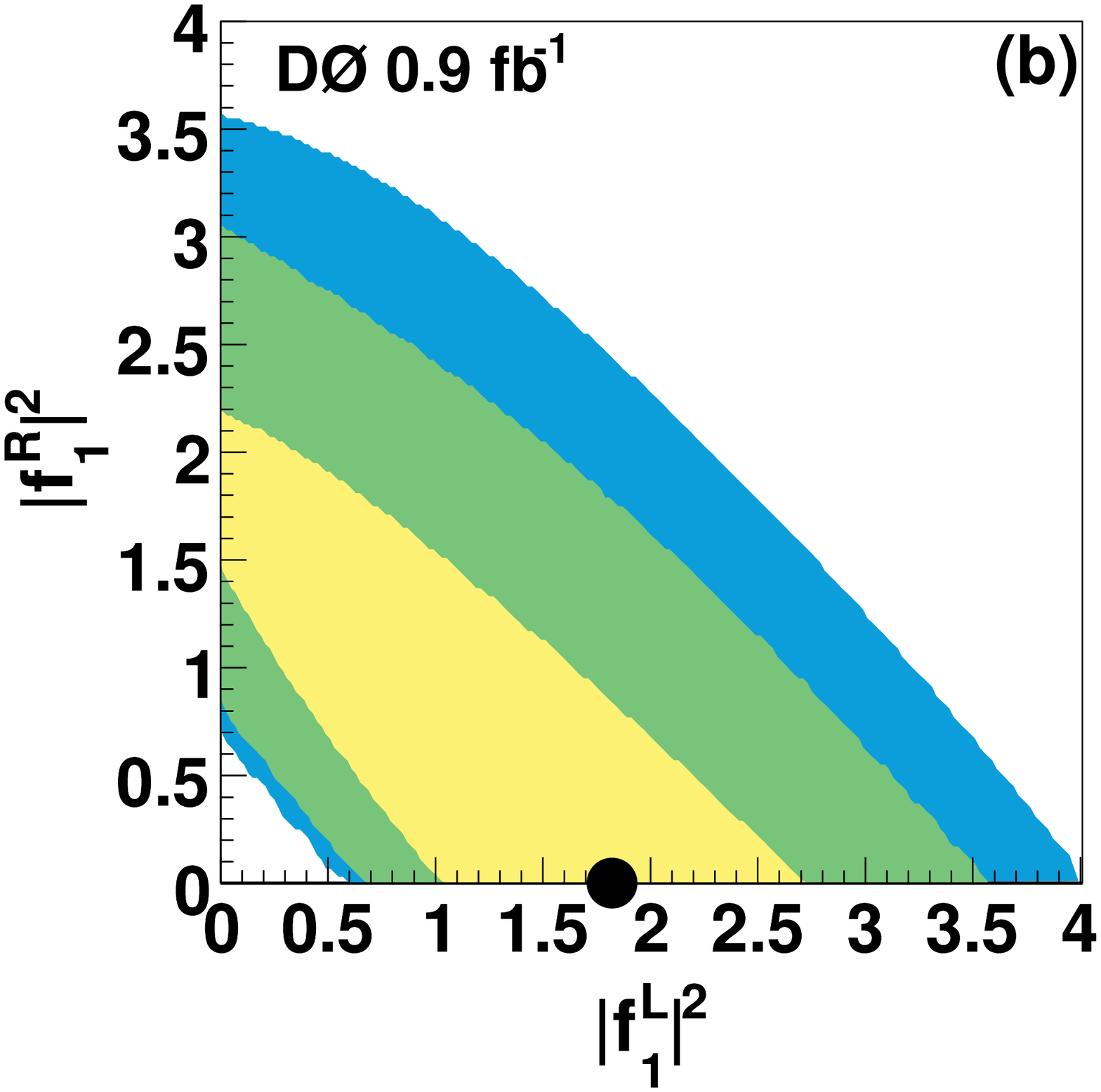}
\includegraphics[width=55mm]{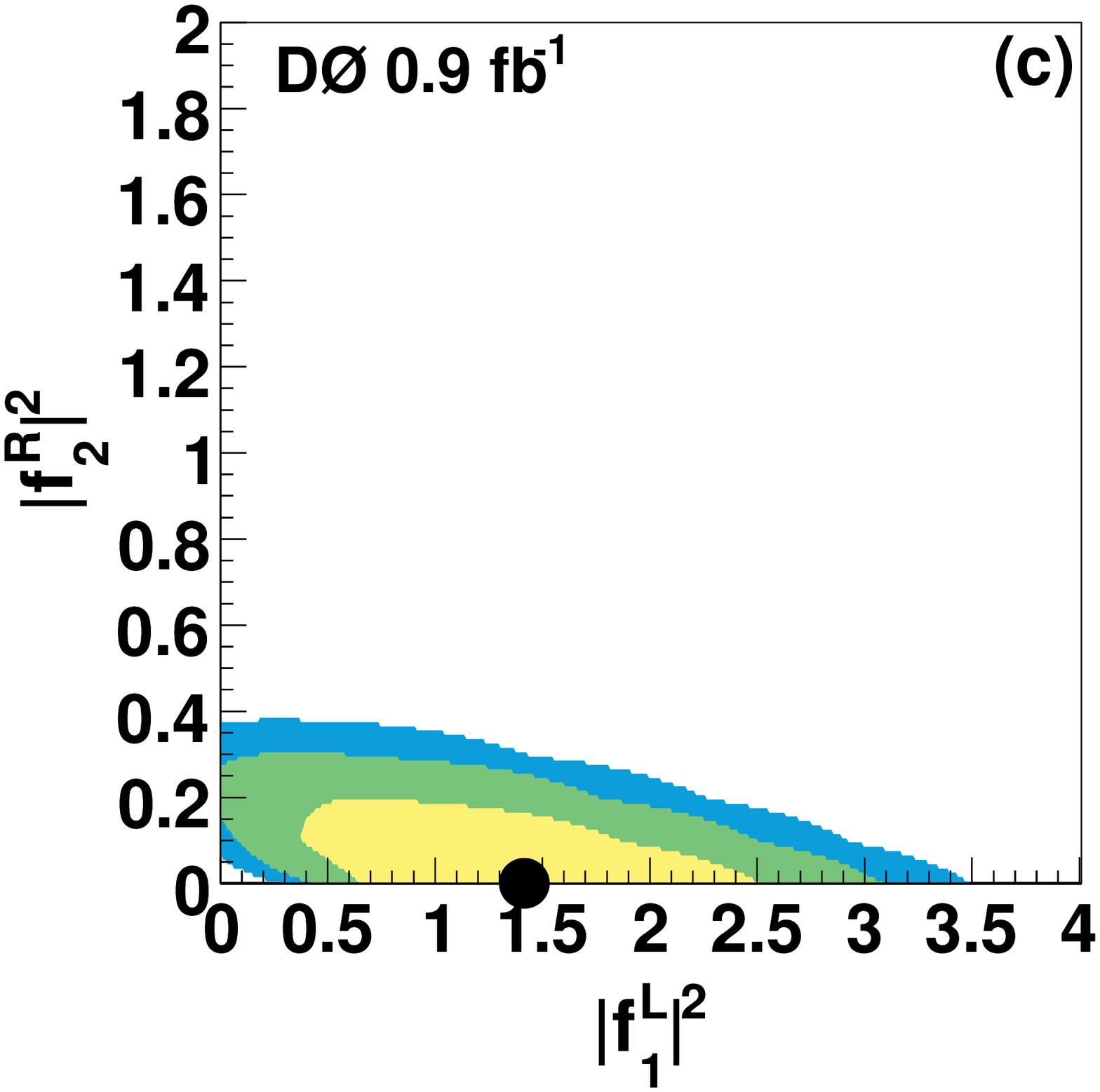}
\caption{Constraints on $tWb$ coupling form factors from single top candidate events.  In each case, the two form factors not plotted are set to zero. } \label{fig:topanomcouplings}
\end{figure*}

In the SM, $f_1^L \approx 1$ and all of the other form factors are zero.  D\O\ searches for non-SM effects by performing a series of measurements in which the boosted decision tree selection is optimized    to discriminate from background signals arising from two possible couplings (the SM $f_1^L$ coupling is always assumed to contribute, and each of the other form factors are tested sequentially)~\cite{D0anomcoupling}.  By fitting the decision tree output distribution in data to expectations from background and the signal model for each form factor, D\O\ measures the contribution of each form factor, and the resulting constraints on the form factors are shown in Fig.~\ref{fig:topanomcouplings}.  At this stage, the results are consistent with purely SM coupling.

\subsection{Search for $H^+ \rightarrow tb$}

If there is a charged Higgs boson with mass greater than that of the top quark, it would be expected to decay preferentially to $tb$, resulting in the same set of final-state particles as in s-channel single-top quark production.  Consequently,  the same selection tools used in the search for the single top can also be used to search for the $H^+$.  If the $H^+$ is a narrow resonance, its contribution will appear as an enhancement in the $Wjj$ mass distribution.  D\O\ has searched for such a resonance in 0.9~fb$^{-1}$ of data, and finds that the data is consistent with purely SM production~\cite{D0hplus1}.  This translates to 95\% C.L.  lower limits on the $H^+$ mass ranging from 184 GeV to 181 GeV as the assumed value of $\tan\beta$ is varied from 25 to 50.  No mass limit can be set for $\tan\beta > 70$, since such models violate the assumption that the $H^+$ has a narrow width.

\subsection{Search for flavor-changing neutral currents in single top production}

Flavor-changing neutral-current (FCNC) couplings would tend to increase the rate of single top quark production, both at the Tevatron and at the HERA $ep$ collider.   An anomalous coupling of the form $tu\gamma$ could lead to top quark  production at HERA, and the H1 and ZEUS experiments have searched for $t \rightarrow Wb$~\cite{Chekanov:2003yt,Aktas:2004ij}.   H1 has updated these results, using 482 pb~$^{-1}$ of data to search for $t\rightarrow Wb$ with the $W$ boson decaying leptonically~\cite{H1FCNC}.    A multivariate discriminant is used to distinguish top quark production from SM backgrounds.  The output distribution of this discriminant is consistent with SM production, providing a limit of $\sigma_{ep\rightarrow etX} < 0.16$ pb at 95\% C.L.  This corresponds to a limit on the $tu\gamma$ magnetic coupling constant of $\kappa_{tu\gamma}< 0.14$, where the cutoff scale $\Lambda$ is taken to be the top quark mass.

CDF searches for anomalous single-top production arising from the FCNC $gtu$ or $gtc$ couplings~\cite{CDFFCNC} (an earlier measurement by D\O\ is described in~\cite{D0FCNC}).  This kind of anomalous production changes both the kinematic distributions and the $b$ jet content in the final state with respect to the SM expectation.  CDF exploits these differences by training a neural network to discriminate between anomalous single top production and background, where background  includes SM single top production.  No evidence of anomalous production is observed, which provides limits on the FCNC coupling constants of 
$\kappa_{gtu}/\Lambda < 0.025 \hbox{TeV}^{-1}$ and $\kappa_{gtc}/\Lambda < 0.105 \hbox{TeV}^{-1}$ at 95\% C.L.

\section{MEASUREMENTS USING $t\bar{t}$  EVENTS}

Most top quark properties are measured using $t\bar{t}$ pairs  produced via the strong interaction.  Not only is the cross section of $\approx 7$~pb for this process higher than that for single top production at the Tevatron, but $t\bar{t}$ events are also more easily distinguished from background.  With two top quarks in the event, several final-state signatures arise from the different decay channels of the two $W$ bosons.  The rarest is when both $W$ bosons decay leptonically (the ``dilepton'' mode), accounting for about 5\% of all $t\bar{t}$ final states.  The mode in which one $W$ decays leptonically and the other hadronically (the ``lepton plus jets'' mode) accounts for 30\%, modes with one or more $\tau$ leptons account for 19\%, and the remaining 45\% of the events have both $W$ bosons decaying hadronically (the ``all jets'' mode).  
The lepton plus jets mode offers the best combination of rate and purity, but it is important to study the other modes as well, since new physics may contribute with different strengths to different final states.

\subsection{Measurement of the $gg$ fusion fraction in $t\bar{t}$ production}

In the SM, we expect 85\% of $t\bar{t}$ production at the Tevatron to arise from  $q\bar{q}$ annihilation, with the remainder from $gg$ fusion.   CDF has measured the $gg$ production fraction $F_{gg}$ using three different analysis techniques~\cite{CDFprodmech}.  The first two of these use events in the lepton plus jets final state.  One analysis compares the number of low-$p_T$ tracks in $t\bar{t}$ candidate events with those in data control samples arising from gluon-mediated (QCD dijets) and quark-mediated ($W$ or $Z$ boson) processes.   Since gluons tend to radiate additional soft gluons more readily than quarks do, gluon-mediated events tend to have more low-$p_T$ tracks.  The second analysis combines kinematic variables into a neural network to discriminate between $q\bar{q}$ and $gg$ initial states (angular variables that are sensitive to the difference in spin between the two initial states are especially useful).  These measurements are combined to yield  $F_{gg} = 0.07^{+0.15}_{-0.07}$ (stat. + syst.).   CDF's third measurement uses events in the dilepton final state, in which the azimuthal angle between the leptons is sensitive to the production mechanism.  This analysis yields  $F_{gg} = 0.53^{+0.35}_{-0.37}\hbox{ (stat.)} ^{+0.07}_{-0.08} \hbox{ (syst.)}$.  

\subsection {Measurement of the $t\bar{t}$ production cross section}

\begin{figure*}[t]
\centering
\includegraphics[width=80mm]{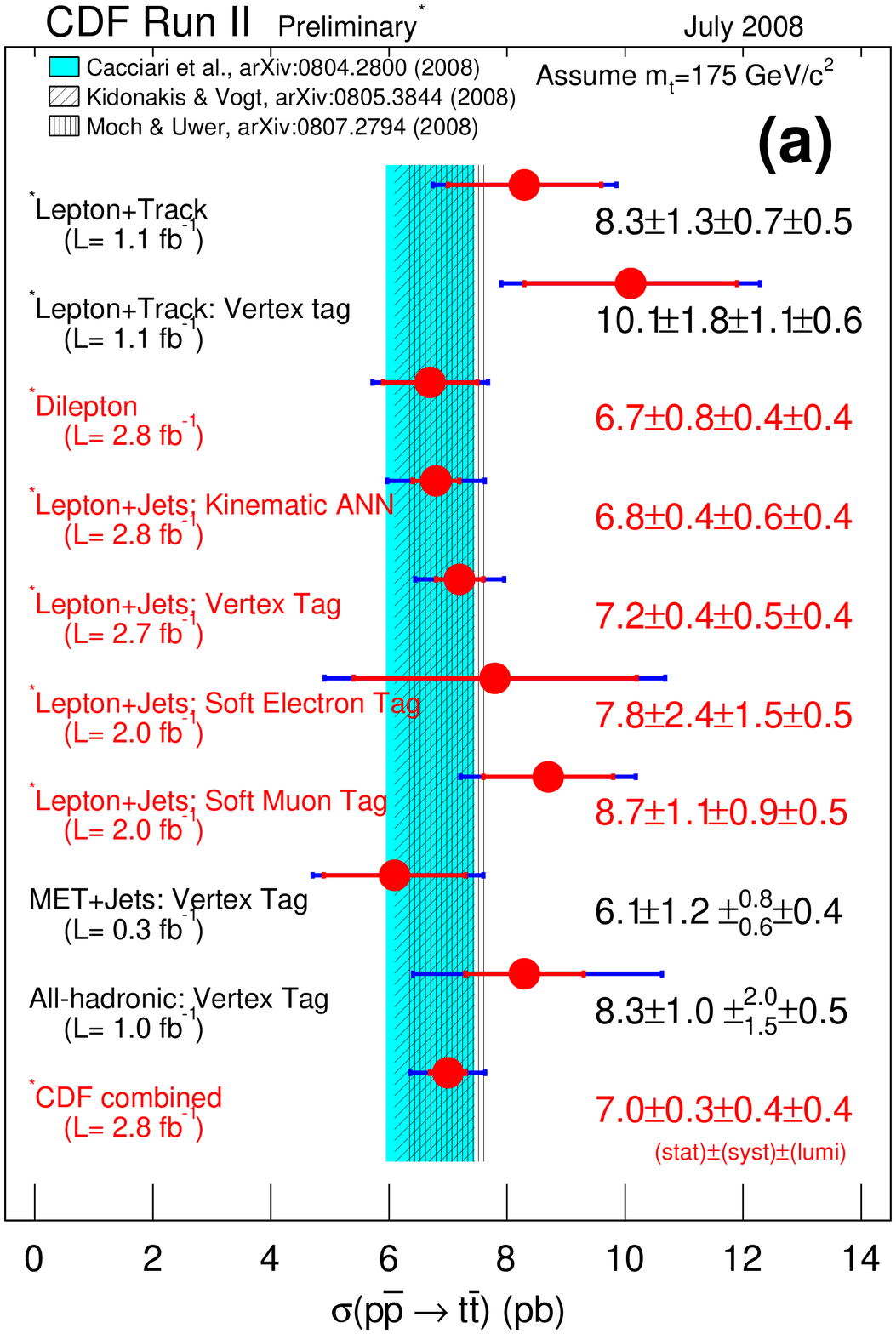}
\includegraphics[width=80mm]{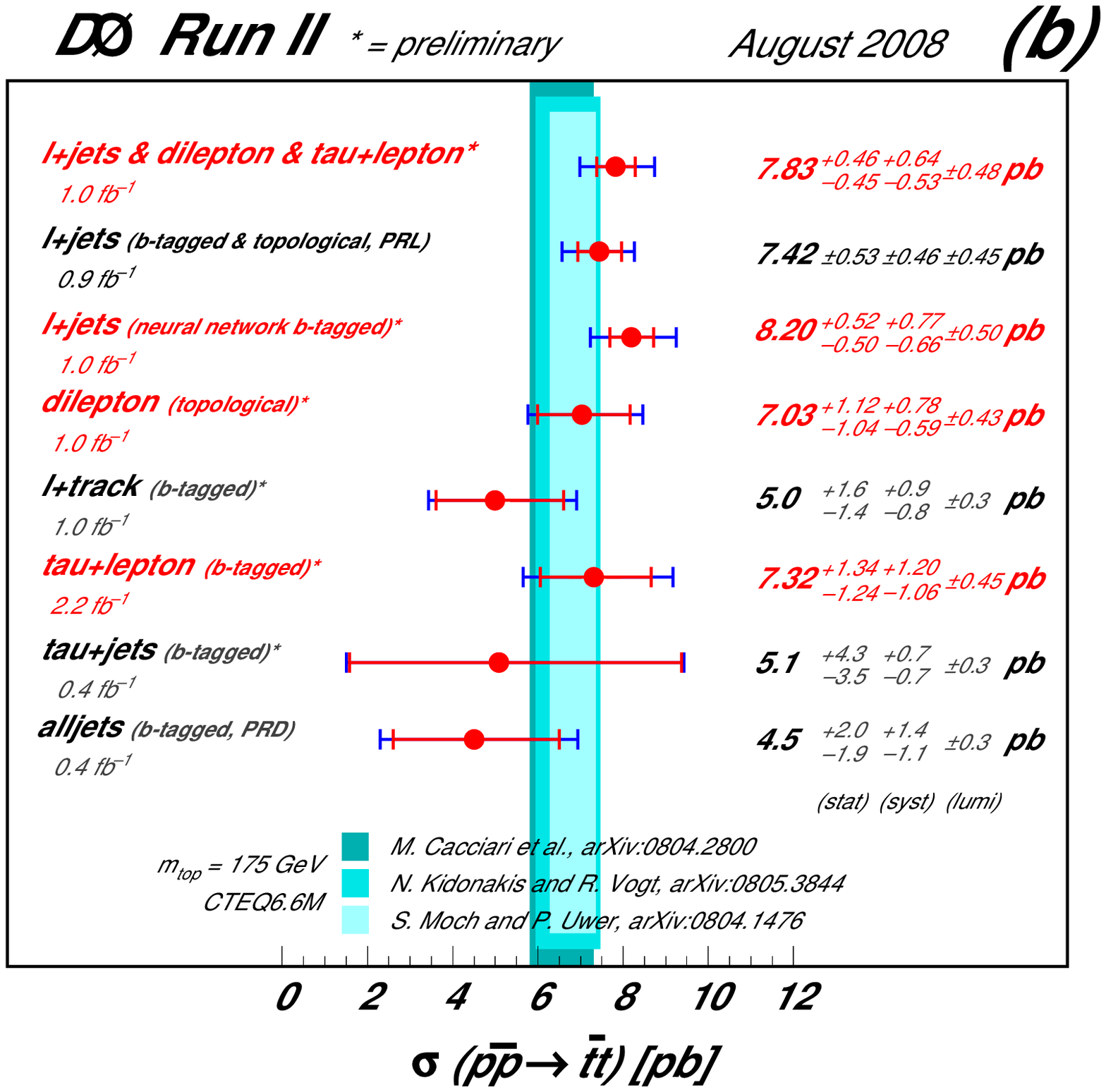}
\caption{Summary of $t\bar{t}$ production cross section measurements from (a) CDF and (b) D\O. } \label{fig:topxs}
\end{figure*}

Measuring the $t\bar{t}$ production cross section is important because a discrepancy between the measured and the calculated values would indicate the presence of new physics either in the production or decay of top quarks.  This measurement requires the detector response to leptons and jets to be well calibrated, and backgrounds well understood.  To maximize sensitivity to new physics, both CDF and D\O\ have measured the production cross section in all final states.  A summary of these measurements is presented in Fig.~\ref{fig:topxs}, which indicates no inconsistency between the final states.  Combining  measurements within each experiment yields $\sigma_{t\bar{t}} = 7.0 \pm 0.3 \hbox{(stat.)} \pm 0.4 \hbox{(syst.)} \pm 0.4  \hbox{(lumi.)}$ pb for CDF~\cite{CDFxscomb} and $\sigma_{t\bar{t}} = 7.8 \pm 0.5 \hbox{(stat.)} \pm 0.6 \hbox{(syst.)} \pm 0.5  \hbox{(lumi.)}$ pb for D\O~\cite{D0xscomb}. In both cases, the total uncertainty is similar to the uncertainty in the theoretical prediction, so that improvements in calculations will be required to provide sensitivity to any new physics resulting from this measurement.

\subsection {Measurement of the top quark mass}

Precision measurement of the top quark mass is of interest both because it is a parameter of the SM and also because it contributes to information about the mass of the Higgs boson.  The latter arises from sensitivity to loop corrections to the $W$ boson mass, which are dominated by contributions from the top quark and Higgs boson. Hence, if both the $W$ boson and top quark masses are measured precisely, the Higgs boson mass can be constrained.  

Assuming SM production and decay, the top quark mass can be estimated by comparing  the measured $t\bar{t}$ production cross section with the theoretical expectation as a function of hypothesized $m_t$.  This has the advantage that the resulting mass is calculated in a well-understood renormalization scheme.  D\O\ uses its cross section measurement to obtain $m_t = 167.8 \pm 5.7\hbox{ (stat. + syst.)}$ GeV or $169.6 \pm 5.4\hbox{ (stat. + syst.)}$ GeV \cite{D0massfromxs}, depending on the theoretical calculation.

\begin{figure*}[t]
\centering
\includegraphics[width=55mm]{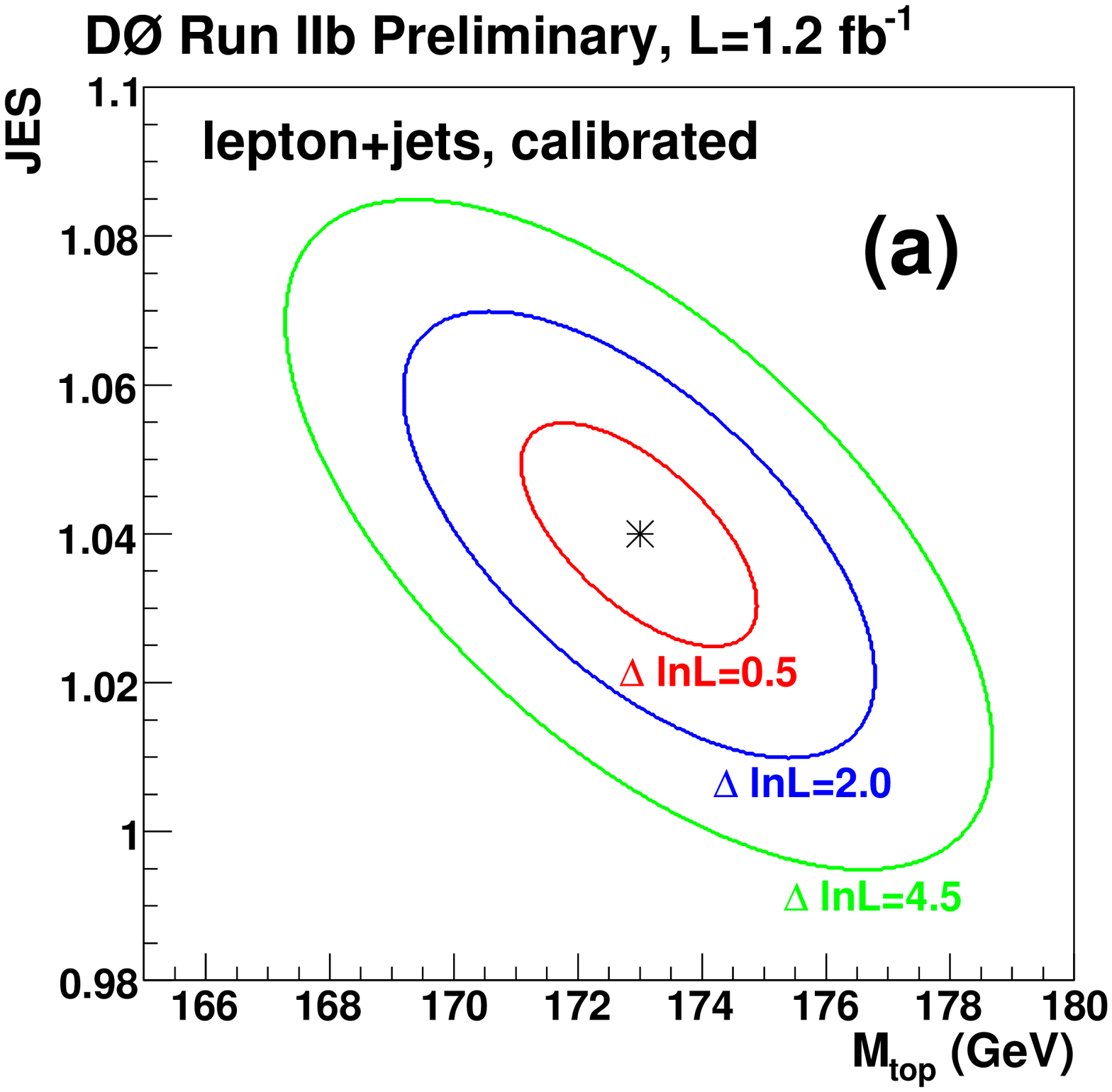}
\includegraphics[width=75mm]{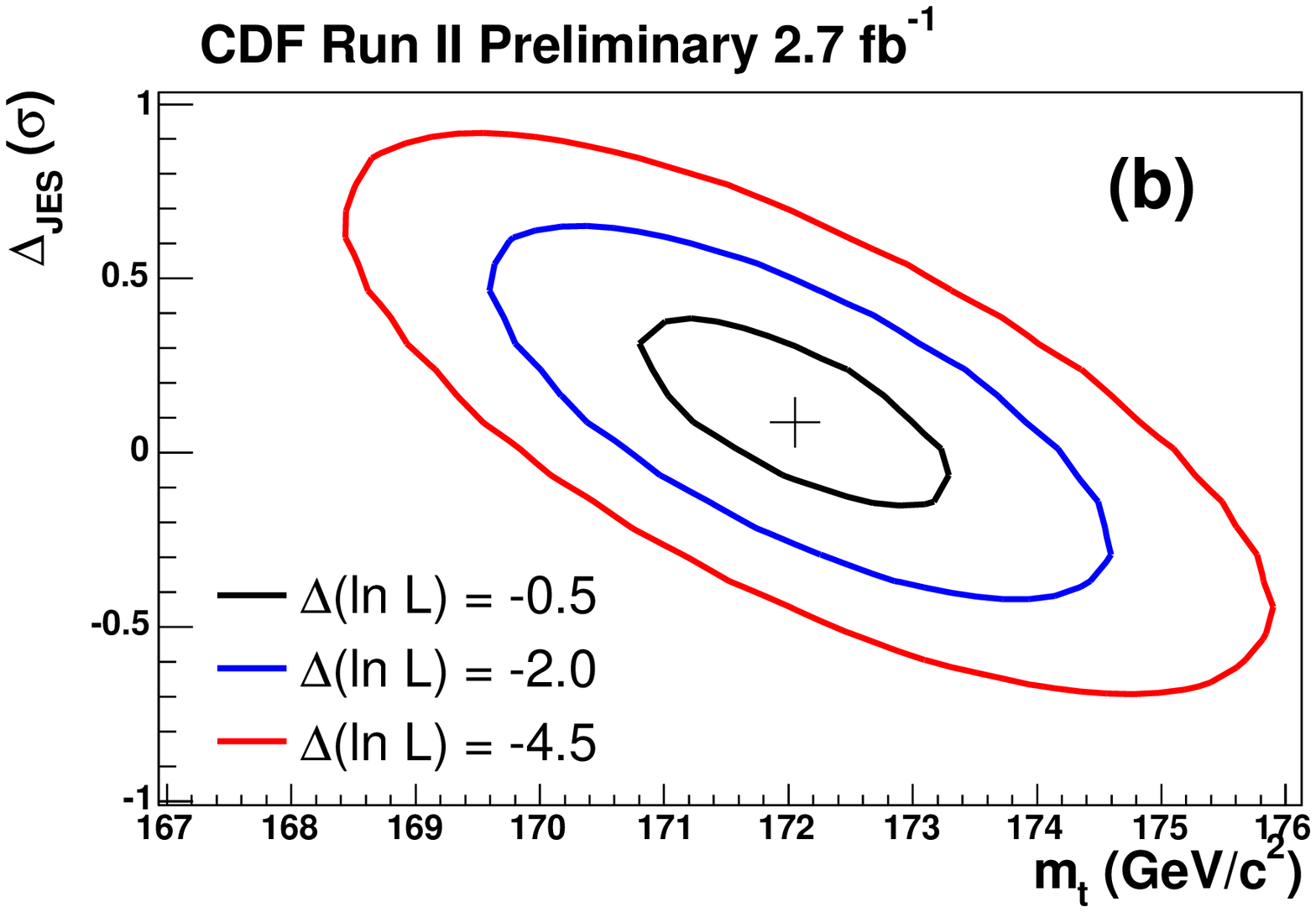}\\
\includegraphics[width=75mm]{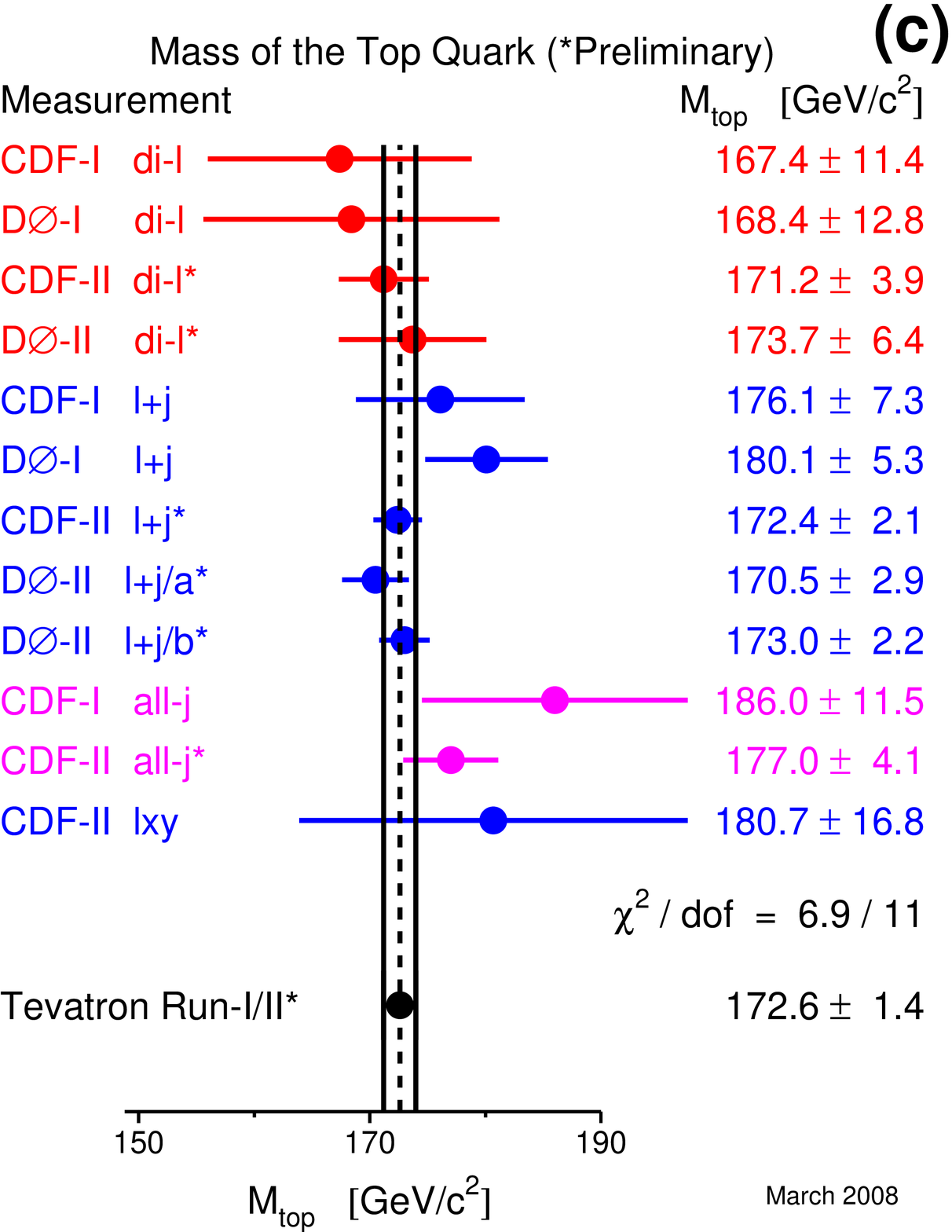}
\caption{Contours in likelihood for the top quark mass vs. jet energy scale (JES) for lepton plus jets events from D\O\ (a) and CDF (b).  Plot (c) provides a summary of all measurements used in the current world-average top quark mass.} \label{fig:topmass}
\end{figure*}

The most accurate measurements of the top quark mass apply the matrix element method to events in the lepton plus jets final state.  Here the matrix element technique is used not only to discriminate signal from background (assigning the most signal-like events the largest weight in the fit), but also to discriminate between different top mass hypotheses.  Approximations in the method (e.g, the  impact of using only the leading-order $q\bar{q}$ annihilation diagram for signal, and considering only the $W$+jet background contribution) are taken into account by applying the method to MC samples of known top mass, and correcting for any observed offsets.  One attractive feature of the lepton plus jets final state is the presence of a hadronically-decaying $W$ boson.  Reconstructing the mass of this boson provides additional information about the jet energy calibration, thereby reducing the systematic uncertainty.  D\O\ and CDF
apply this technique to 2.2 and 2.7 fb$^{-1}$ of data to find, respectively,  $m_t = 172.2 \pm 1.0 \hbox{ (stat.)}\pm 1.4\hbox{ (syst.)}$
GeV \cite{D0ljetstopmass} and $172.2 \pm 1.0\hbox{ (stat.)} \pm 1.3\hbox{ (syst.)}$ GeV \cite{CDFljetstopmass}.  The likelihood for the data to be consistent with various top quark mass and jet energy calibration hypotheses is shown in Fig.~\ref{fig:topmass} (only a subset of the D\O\ data is presented in the plot).  

While the above are the most precise single mass measurements, it is important to measure the top quark mass in a variety of final states,  since the presence of new physics may lead to inconsistencies.  To that end, both D\O\ and CDF have applied the matrix element method to events in the dilepton final state.
With 2.0 fb$^{-1}$ of data, CDF measures $m_t = 171.2 \pm 2.7\hbox{ (stat.)} \pm 2.9\hbox{ (syst.)}$ GeV \cite{CDFdileptopmass} while D\O\ uses a sample of up to 2.8 fb$^{-1}$ to measure $m_t = 174.4 \pm 3.2 \hbox{ (stat.)}\pm 2.1\hbox{ (syst.)}$ GeV \cite{D0dileptopmass}, consistent with the measurements in the lepton plus jets channel.

CDF has also performed a measurement of the top quark mass using the all jets channel ~\cite{CDFalljetsmass}.  Though it has the largest $t\bar{t}$ branching ratio, this mode is difficult to extract from the large multijet background.  CDF combines $b$ jet identification with a kinematic neural network selection to this channel, and extracts the top quark mass from a full kinematic reconstruction of the final state.  As with the lepton plus jets mode, hadronic $W$ decays provide an {\it in situ} measurement of the jet energy calibration.  The result of the measurement is  $m_t = 176.9 \pm 3.8 \pm 1.7$ GeV.

In addition, CDF has measured the top quark mass using the distributions in lepton $p_T$ and $b$ decay length~\cite{CDFjesindepmass}.   These variables are independent of the jet energy calibration, and so the systematic uncertainties in this measurement are largely uncorrelated with those in the other measurements.  The result is $m_t = 175.3 \pm 6.2\hbox{ (stat.)} \pm 3.0\hbox{ (syst.)}$ GeV.  While this is currently less precise than the matrix element results, such techniques should be valuable at the LHC.
 
The above (and previous) measurements are combined by the Tevatron Electroweak Working Group to obtain a world-average top quark mass of $172.4 \pm 0.7\hbox{ (stat.)} \pm 1.0\hbox{ (syst.)}$ GeV, with a precision of 0.7\% \cite{WAtopmass}.  Using this value as one of the inputs to a global fit of electroweak parameters of the SM, the LEP Electroweak Working Group finds that the Higgs boson mass is constrained to be less than 154 GeV at 95\% C.L.~\cite{LEPEWWG}.  If the direct search from LEP is taken as a prior probability, this upper limit increases to 185 GeV. 

\subsection{Forward-backward charge asymmetry}

The distribution of the rapidity difference $\Delta y \equiv y_t - y_{\bar{t}}$ is expected to have a small deviation from symmetry about $\Delta y=0$.  This asymmetry arises from interference effects between production diagrams at NLO, and can be quantified by:
\begin{equation}
A_{\rm fb} \equiv {N_{\rm f} - N_{\rm b} \over  N_{\rm f} + N_{\rm b}} 
\end{equation}
where $N_{\rm f}$ and $N_{\rm b}$ are the numbers of events with positive and negative values of $\Delta y$.  In the SM, $A_{\rm fb}$ is expected to be of the order 5-10\%, but new physics could significantly increase this value.

Both CDF and D\O\ measure $A_{\rm fb}$ using lepton plus jets events.  Event selection and reconstruction can alter the distribution of $\Delta y$, so that the apparent value of $A_{\rm fb}$ differs from its true value.  CDF and D\O\ take different approaches to addressing this issue, with CDF choosing to unfold its measured asymmetry under the assumption that the production is according to the SM, while D\O\ reports its measured asymmetry and provides a prescription for modeling the distortion of any true $\Delta y$ distribution.  This difference means that the results reported by the two experiments cannot be compared directly.  

CDF performs two measurements using 1.9 fb$^{-1}$ of data, with the asymmetry determined in the $t\bar{t}$ and $p\bar{p}$ rest frames.  The former analysis yields $A_{\rm fb} = 24 \pm 13 \hbox{(stat.)} \pm 4\hbox {(syst.)}$\% and the latter $A_{\rm fb} = 17 \pm 7 \hbox {(stat.)} \pm 4\hbox {(syst.)}$\%.  
D\O\ measures $A_{\rm fb}$ using 0.9 fb$^{\-1}$ of data, and finds  $A_{\rm fb} = 12 \pm 8 \hbox {(stat.)} \pm 1\hbox {(syst.)}$\%.

\subsection{Searches for $t\bar{t}$ resonances}

Both CDF and D\O\ have searched for heavy particles that decay into $t\bar{t}$ by reconstructing the invariant mass of the $t\bar{t}$ system $M_{t\bar{t}}$.  A resonance would appear as a localized excess in the $M_{t\bar{t}}$ distribution.  This has been studied at CDF and D\O\ using 1.9 and 2.1 fb$^{-1}$ of data, respectively~\cite{CDFmtt, D0mtt}.  Neither experiment finds evidence for resonant $t\bar{t}$ production, and they  set limits on such resonances.  CDF expresses the limits as restrictions on heavy gluons as a function of the gluon mass and coupling to the top quark.  D\O\ expresses the limits in terms of a $Z^\prime$ resonance, and finds that if a $Z^\prime$ exists, it must have a mass $> 760$ GeV at 95\% C.L.

\subsection{Search for a $t^\prime$ quark}

Although the SM has only three generations of fermions, there may in fact be additional generations of massive fermions.  CDF searches for a fourth-generation analogue to the top quark (the $t^\prime$ quark) in 2.8 fb$^{-1}$ of data, assuming that this quark is strongly pair-produced and decays to $Wq$~\cite{CDFtprime}.  As a consequence, the experimental signature for $t^\prime\bar{t^\prime}$ production is similar to that of $t\bar{t}$ production, with the exception that $t^\prime$ events would have more energetic jets and leptons and a larger reconstructed quark mass.  Comparing the two-dimensional distribution of reconstructed mass versus the sum of the jet and lepton $p_T$ values in lepton plus jet events, CDF finds no evidence for a deviation from the SM expectation, and sets a lower limit on the mass of a $t^\prime$ of 311 GeV  at 95\% C.L..

\subsection{Measurement of the $W$ boson helicity}

For the $V-A$ coupling of the SM , the $W$ boson arising from top quark decay should have helicity 0 70\% of the time, helicity -1 30\% of the time, and a only a negligible contribution from helicity +1. Departures from $V-A$ for the $tWb$ coupling alter these fractions.   Both CDF and D\O\ have measured the $W$ helicity fractions by reconstructing the angle $\theta^*$ between the up-type fermion and top quark directions in the $W$ boson rest frame.  The CDF measurement is based on two analyses of lepton plus jets events: one in which the distribution of 
$\cos\theta^*$ is unfolded and compared directly to the theoretical prediction~\cite{CDF_whel_unfold}, and the other in which MC templates are used to extract the underlying helicity fractions from the observed distribution~\cite{CDF_whel_templ}.
D\O's measurement uses the lepton plus jets and $e\mu$ channels,  taking advantage of information from both $W$ bosons in each event in a template fit to the data.~\cite{D0_whel}.    All the analyses  simultaneously fit  the fractions $f_0$ and $f_+$ of $W$ bosons with helicity 0 and +1.  Combining its two analysis~\cite{CDF_whel_comb}, CDF finds $f_0 = 0.66 \pm 0.16$ (stat. + syst.) and $f_+ = -0.03 \pm 0.07$ (stat. + syst.), consistent with the SM.  D\O\ finds $f_0 = 0.49 \pm 0.10 \hbox{(stat.)} \pm 0.08 \hbox{(syst.)}$ and $f_0 = 0.11 \pm 0.05 \hbox{(stat.)} \pm 0.05 \hbox{(syst.)}$.  D\O's measurement has a $p$-value of 0.23 for being consistent with the SM.

\subsection{Measurement of the top quark decay branching fractions}

The prediction that the top quark should almost always decay to $Wb$ can be tested using the number of $b$-tagged $t\bar{t}$ candidate events.  D\O\ uses the yield of $t\bar{t}$ events with 0, 1, and 2 $b$-tagged jets in 0.9 fb$^{-1}$ of lepton plus jets data to simultaneously measure the $t\bar{t}$ production cross section and the ratio
\begin{equation}
R\equiv {{\cal B} \rightarrow Wb \over {\cal B} \rightarrow Wq} = { |V_{tb}|^2 \over |V_{td}|^2 + |V_{ts}|^2 + |V_{tb}|^2}.
\end{equation}
The results are: $\sigma_{t\bar{t}} = 8.18^{+0.90}_{-0.84} \hbox{ (stat.+syst.)} \pm 0.50 \hbox{ (lumi.)}$ pb
and $R = 0.97^{+0.09}_{-0.08} $ (stat. + syst)~\cite{D0r}.

CDF reports a measurement of the yield of events with two $b$-tagged jets in 1.9 fb$^{-1}$ of lepton plus jets data~\cite{CDFinvis}.  Top quark decays to any final state other than $Wb$ would result in a deficit of doubly-tagged events relative to the SM prediction.  CDF measures the yield in exclusive jet multiplicities of 2, 3, and 4, and in the inclusive multiplicity of $\ge 5$ jets.  In each sample the yield is consistent with the SM prediction.  Using this information, CDF sets 95\% C.L. upper limits on non-SM branching ratios, finding ${\cal B}(t\rightarrow Zc) < 13\%$ and ${\cal B}(t \rightarrow \hbox{invisible}) < 9\%$, where invisible refers to any final state that would not provide a $b$-tagged jet.
  
\subsection{Search for FCNC in top quark decay}

In addition to the searches for FCNC in single top quark production described above, CDF has also searched for the FCNC decay $t \rightarrow Zq$ in 1.9 fb$^{-1}$ of data~\cite{CDF_FCNC_decay} comprised of events with two leptons of opposite charge and mass consistent with the $Z$ boson and containing four or more jets.  This is the signature that arises from $t\bar{t} \rightarrow Z(\rightarrow \ell^+\ell^-)qW(\rightarrow q\bar{q}^\prime)b$.  Since there are no neutrinos in this final state, the event can be fully reconstructed.  Using a template to fit the distribution of the kinematic $\chi^2$ for each event to be consistent with the above decay hypothesis, CDF extracts the fraction of FCNC top quark decays in the data sample.  No evidence for FCNC is found,  and ${\cal B}(t \rightarrow Zq)$  is constrained to be $ < 3.7\%$ at 95\% C.L.

\subsection{Search for $t \rightarrow H^+b$}

\begin{figure*}[t]
\centering
\includegraphics[width=80mm]{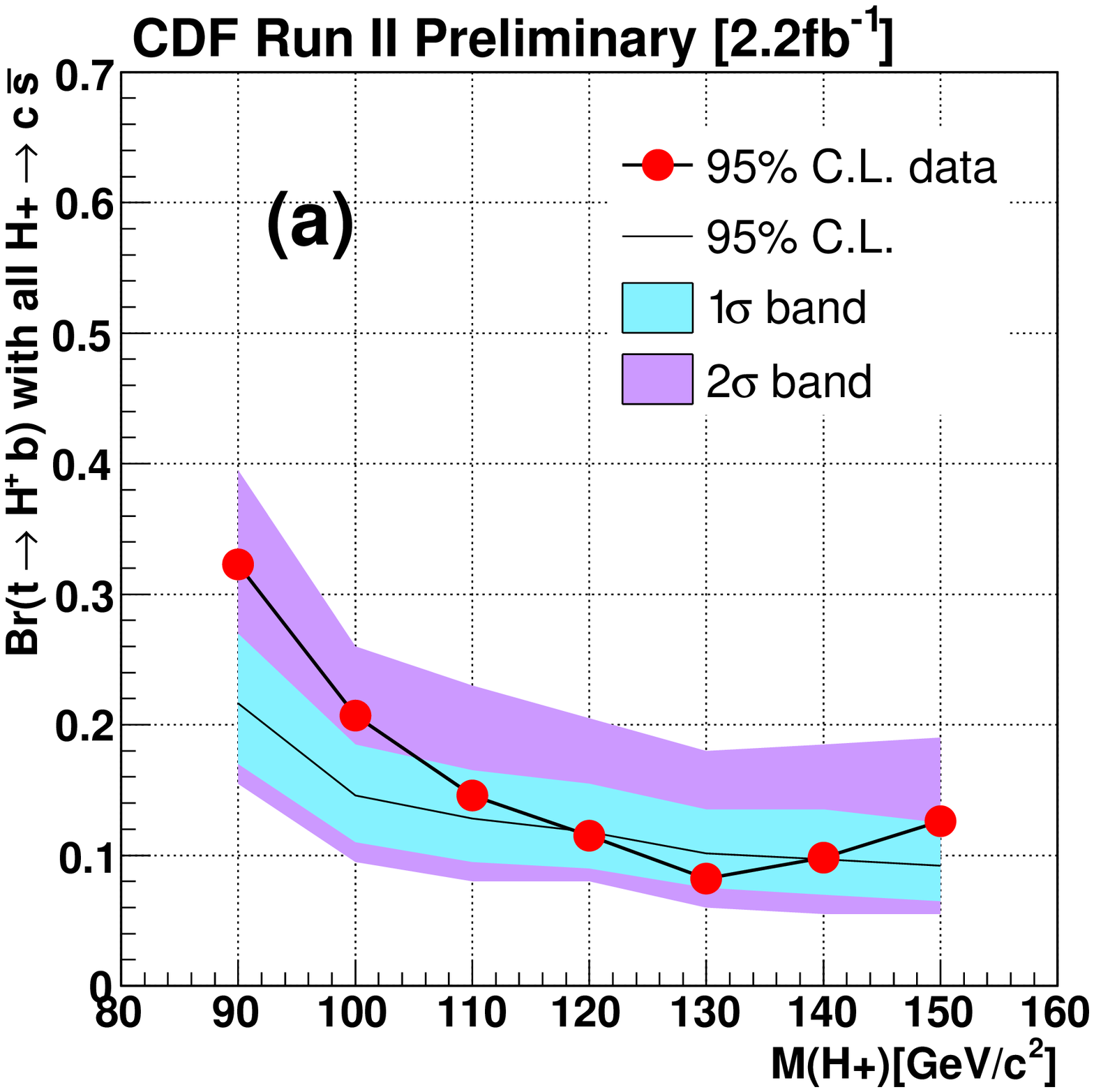}
\includegraphics[width=80mm]{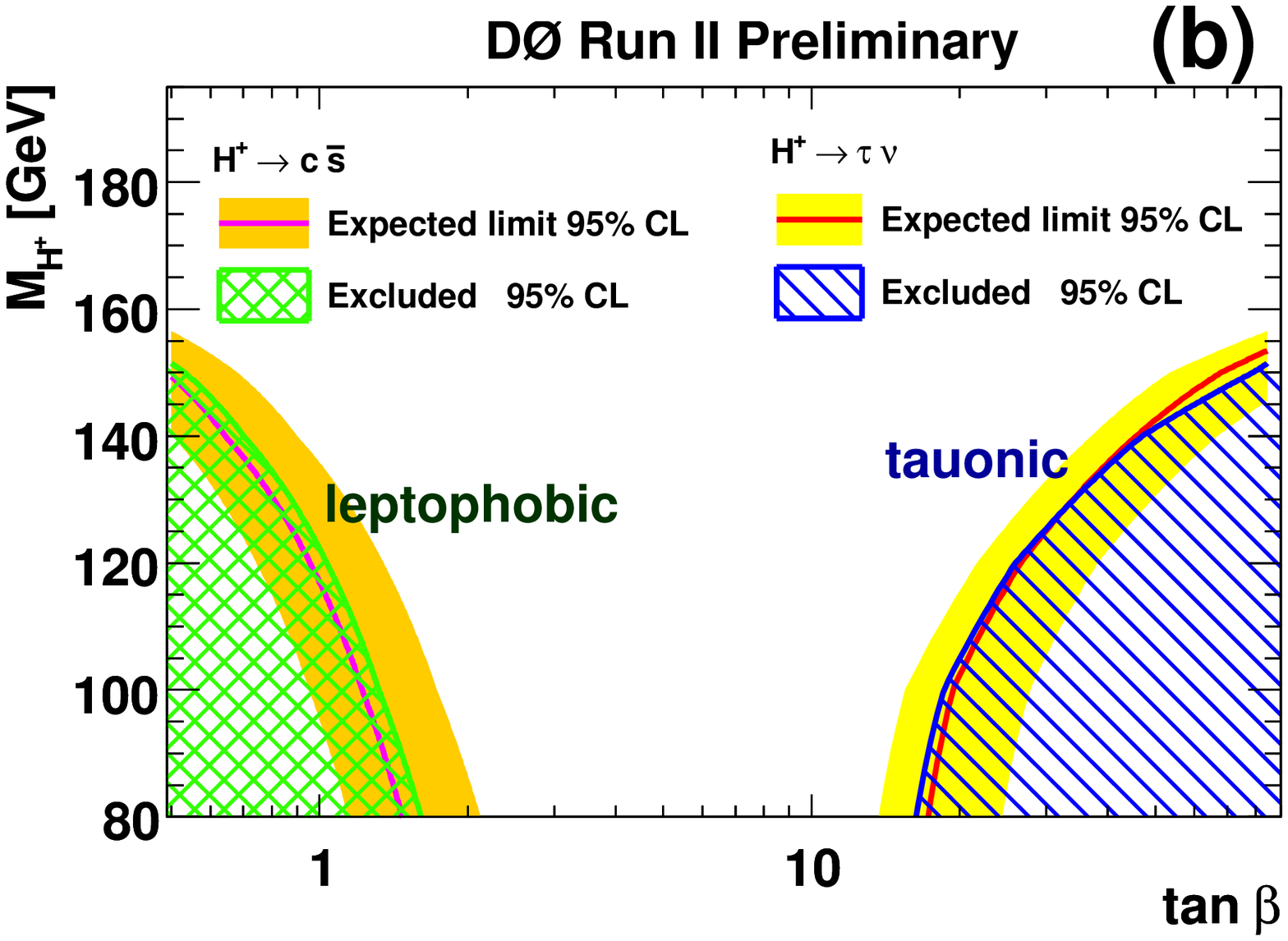}
\caption{Limits on the charged Higgs boson arising from searches for $t \rightarrow H^+b$ by the (a) CDF and (b) D\O\ collaborations.  The CDF limits assume that ${\cal B}(H^+ \rightarrow c\bar{s}) = 1$, while the D\O\ limits are given in the context of the MSSM.} \label{fig:hpluslimits}
\end{figure*}

As a complement to the search for $H^+ \rightarrow tb$ production, both CDF and D\O\ search for the decay $t \rightarrow H^+b$ to explore the scenario where $m_{H^+} < m_t$.   CDF's analysis~\cite{CDFhplusdecay} is based on the kinematics of events in the lepton plus jets decay mode with two $b$-tagged jets.  This search is sensitive to the decay $H^+\rightarrow c\bar{s}$, which is dominant in the minimal supersymmetric standard model (MSSM) at low $\tan\beta$ (the analysis assumes that ${\cal B}(H^+ \rightarrow c\bar{s}) = 1$).  CDF searches for an $H^+$ resonance in the  invariant mass of the two leading non-tagged jets using 2.2 fb$^{-1}$ of data (in some cases in events with more than four jets, one of the``extra jets" is added to the leading jets under the assumption that it arose from final-state radiation).   CDF observes no such resonance, and sets limits in the  ${\cal B} (t \rightarrow H^+b)$ vs. $m_H^+$ plane, as shown in Fig.~\ref{fig:hpluslimits}. 

D\O's analysis~\cite{D0xscomb} utilizes a different strategy.  In 1.0 fb$^{-1}$ of data,  the event yields in the final states $e+$jets, $\mu$+jets, $ee$, $e\mu$, $\mu\mu$, $e\tau$ and $\mu\tau$ channels are compared to the expectation from the SM.   These yields should change if there are $H^+$ decays to $c\bar{s}$  or to $\tau\nu$ (dominant at large $\tan\beta$ in the MSSM).  D\O\ observes no significant deviation in the yield in any of the final states, and sets limits in the $m_{H^+}$ vs. $\tan\beta$ plane, as shown in Fig.~\ref{fig:hpluslimits}.

\section{TOP QUARK PHYSICS AT THE LARGE HADRON COLLIDER}

While to date only the Fermilab Tevatron has had sufficient energy to produce top quarks, the Large Hadron Collider will be coming online soon.  With its 14 TeV center-of-mass energy, the LHC will have a
$t\bar{t}$ production cross section more than two orders of magnitude greater than that at the Tevatron.
The resulting abundance of statistics will allow substantial improvements in the precision of many of the measurements discussed above.  With an initial data sample of 10 fb$^{-1}$ (about a year of running at low luminosity), the top quark mass can be measured with an uncertainty of $\approx1$ GeV, FCNC can be probed down to the 10$^{-3}$ or 10$^{-4}$ level, and the $W$ boson helicity fractions can be measured to within  
1-2\%~\cite{LHCtopprediction}.

\section{CONCLUSION}

The increasing sample of $t\bar{t}$ events at the Tevatron has enabled the top quark to be probed in many ways at unprecedented levels of precision, as highlighted by the measurement of the top quark mass with a precision of 0.7\%.  Several other measurements of the top quark's production and decay properties, as well as searches for new particles that have signatures similar to that of the top quark, have been performed.  Thus far, none of these measurements indicate any departure from the SM, and
have accordingly constrained models for new phenomena.  Many of these measurements will be repeated and provide dramatic increases in precision in the coming years as large samples from the LHC become available.

% If you have acknowledgments, this puts in the proper section head.
%\begin{acknowledgments}
%
%I would like to thank the conveners of the D\O\ and CDF top quark physics groups 
%
%\end{acknowledgments}

\end{document}